\documentclass[reprint,amssymb,amsmath,aps,prl]{revtex4-1}

\usepackage{graphicx}
\usepackage{amssymb}
\usepackage{epstopdf}
\usepackage{upgreek}
\usepackage{dcolumn}
\usepackage{bm}
\usepackage{gensymb}
\usepackage{braket}
\usepackage{amsmath}
\usepackage{float}

\expandafter\ifx\csname package@font\endcsname\relax\else
 \expandafter\expandafter
 \expandafter\usepackage
 \expandafter\expandafter
 \expandafter{\csname package@font\endcsname}%
\fi

\newcommand{\be}{\begin{eqnarray}}
\newcommand{\ee}{\end{eqnarray}}
\newcommand{\bfig}{\begin{figure}}
\newcommand{\efig}{\end{figure}}

\begin{document}

\title{Breaking Lorentz reciprocity with frequency conversion and delay}
\author{Eric I. Rosenthal}
\thanks{These authors contributed equally to this work.\\ eric.rosenthal@colorado.edu \\ benjamin.chapman@colorado.edu}
\affiliation{JILA, National Institute of Standards and Technology and the University of Colorado, Boulder, Colorado 80309, USA}
\affiliation{Department of Physics, University of Colorado, Boulder, Colorado 80309, USA}
\author{Benjamin J. Chapman}
\thanks{These authors contributed equally to this work.\\ eric.rosenthal@colorado.edu \\ benjamin.chapman@colorado.edu}
\affiliation{JILA, National Institute of Standards and Technology and the University of Colorado, Boulder, Colorado 80309, USA}
\affiliation{Department of Physics, University of Colorado, Boulder, Colorado 80309, USA}
\author{Andrew P. Higginbotham}
\affiliation{JILA, National Institute of Standards and Technology and the University of Colorado, Boulder, Colorado 80309, USA}
\affiliation{Department of Physics, University of Colorado, Boulder, Colorado 80309, USA}
\author{Joseph Kerckhoff}
\altaffiliation{Current address: HRL Laboratories, LLC, Malibu, CA 90265, USA}
\affiliation{JILA, National Institute of Standards and Technology and the University of Colorado, Boulder, Colorado 80309, USA}
\affiliation{Department of Physics, University of Colorado, Boulder, Colorado 80309, USA}
\author{K.~W. Lehnert}
\affiliation{JILA, National Institute of Standards and Technology and the University of Colorado, Boulder, Colorado 80309, USA}
\affiliation{Department of Physics, University of Colorado, Boulder, Colorado 80309, USA}
\date{\today}

\begin{abstract}

We introduce a method for breaking Lorentz reciprocity based upon the non-commutation of frequency conversion and delay. The method requires no magnetic materials or resonant physics, allowing for the design of scalable and broadband non-reciprocal circuits.  With this approach, two types of gyrators --- universal building blocks for linear, non-reciprocal circuits --- are constructed. 
Using one of these gyrators, we create a circulator with $>15~\mathrm{dB}$ of isolation across the $5-9~\mathrm{GHz}$ band.
Our designs may be readily extended to any platform with suitable frequency conversion elements, including semiconducting devices 
for telecommunication or an on-chip superconducting implementation for quantum information processing.
\end{abstract}

\maketitle

Lorentz reciprocity follows from Maxwell's equations, and places strong physical constraints on the operation of electromagnetic devices and networks~\cite{lorentz:1896}.
In the case where propagating fields, entering through ports as guided modes, are scattered by a closed network, Lorentz reciprocity implies that the scattering between a pair of ports is invariant upon exchange of the source and detector~\cite{pozar:2011,jalas:2013}. 
In other words, fields flow backward through the network as easily as they flow forward. When directional scattering is required, such as the ubiquitous case of unidirectional information flow in a communication network, the reciprocity theorem must be broken by violating one of its assumptions.

Typical non-reciprocal elements such as microwave circulators and optical isolators rely on ferromagnetic effects, which are odd under time-reversal, to break Lorentz reciprocity.
This approach, however, is incompatible with some desirable chip-based technologies.
For instance, ferrite circulators cannot be integrated with superconducting qubits and circuits, and Faraday isolators cannot be miniaturized for integration with on-chip photonics.
A broad experimental effort has therefore emerged to develop alternative non-reciprocal devices, including approaches based on: nonlinear materials~\cite{mahmoud:2015,coulais:2017}, quantum Hall physics~\cite{viola:2014,anderson:2016,mahoney:2017,mahoney:2017b}, and active modulation~\cite{anderson:1965,anderson:1966,yu:2009,kamal:2011,lira:2012,kerckhoff:2015,metelmann:2015,reiskarimian:2016,biedka:2017,metelmann:2017,abdo:2017,fang:2017,chapman:2017b}. 

\begin{figure*}[htb]
\begin{center}
\includegraphics[width=1.0\linewidth]{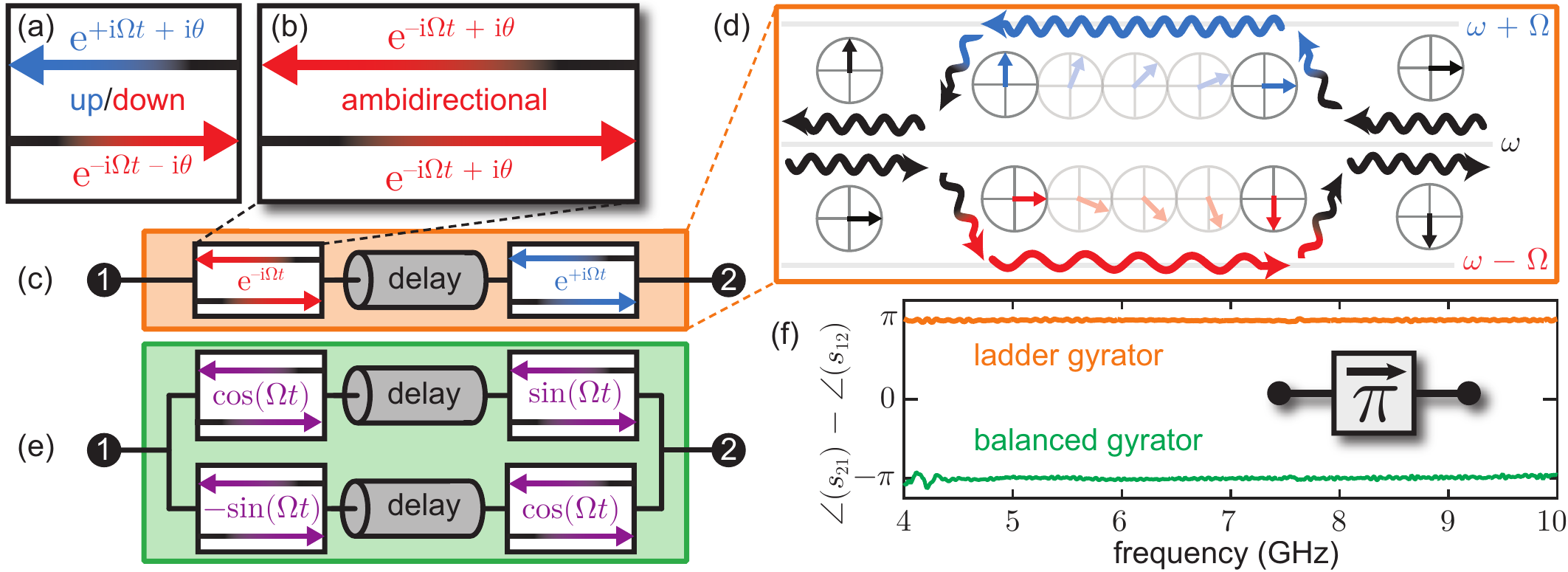}
\caption {
(a) Signal flow graph of an up/down converter used in non-reciprocal circuits~\cite{kamal:2011,abdo:2017}. Signals incident on the left port are down-converted while those incident on the right port are up-converted.
(b) Signal flow graph of an ambidirectional modulator, configured as a down-converter.
Signals incident on either port 
are down-converted.
(c) Schematic of a `ladder gyrator', constructed from two frequency converters concatenated around a delay.
(d) Frequency-time diagram of the ladder gyrator.
Oscillating arrows indicate directed signal flow, and circled arrows indicate the signal's phase in a frame rotating at the input frequency. Signals acquire a non-reciprocal dynamical phase.
(e) Schematic representation of the `balanced gyrator', which is constructed using a `symmetric' frequency converter. Operations $\cos(\Omega t)$ and $\pm \sin(\Omega t)$ convert an input signal into both its upper and lower sideband. 
(f) Measured forward-reverse phase difference, $\angle s_{21} - \angle s_{12}$, as a function of frequency for both gyrators.  For visual clarity, the balanced gyrator trace has been translated down by $2\pi$. Performance is broadband, in the sense that the bandwidth is comparable to the operation frequency. 
}
\label{fig:fig1}
\end{center}
\end{figure*}

Many active circuits realize non-reciprocity with parametric coupling between resonant modes~\cite{estep:2014,ranzani:2015,sliwa:2015,ruesink:2016,lecocq:2017,bernier:2017,peterson:2017,barzanjeh:2017,fang:2017}. The parametric interaction creates a frequency conversion process, illustrated in Fig.~\ref{fig:fig1}a. Time-dependence of a system parameter is the fundamental source of the non-reciprocity: the phase of parametric modulation provides a gauge freedom~\cite{fang:2012b,fang:2013,tzuang:2014} and imprints a non-reciprocal phase shift on the frequency-converted signals. However, such approaches have been limited in bandwidth to a small fraction of the operating frequency, constrained by the linewidths of the coupled resonant modes. 

In this Letter, we propose and demonstrate an alternative method for breaking reciprocity, based upon the non-commutation of frequency conversion and delay. This approach requires no resonant physics, and allows for the design of broadband circuits, in which the bandwidth is comparable to the operating frequency. We construct two broadband microwave gyrators using distinct frequency-conversion elements. A gyrator is a two-port network which imparts no phase shift to forward-propagating signals and a $\pi$ phase shift to reverse-propagating signals~\cite{pozar:2011}; it is also a fundamental building block for creating any passive, linear non-reciprocal circuit~\cite{tellegen:1948}. To demonstrate the utility of a broadband non-reciprocal building block, we use one of our gyrators to construct a circulator with over 15 dB of isolation across the 5 -- 9 GHz band. 

\textit{Theory of operation.---}
The basic source of non-reciprocity in the convert-delay scheme is an ambidirectional modulator followed by a non-commuting time delay. 
A frequency converter is ambidirectional if it applies the same modulation to a signal irrespective of which port it entered from, a property of the converters of Fig.~\ref{fig:fig1}b~and~e, but not Fig.~\ref{fig:fig1}a.
To create the first gyrator, a frequency up-converter and a down-converter are concatenated around a delay (Fig.~\ref{fig:fig1}c~and~d).
We call this device the ladder gyrator, as incident signals are converted up or down on a ladder of frequencies, `shelved' on a rung of this ladder while they accumulate a non-reciprocal dynamical phase, and are then converted back to their original frequency. 

The transfer operator $E$ for a down-converter translates an itinerant mode $\mathbf{a}_\omega$ by a frequency $\Omega < \omega$ such that $E \mathbf{a}_{\omega} = \mathbf{a}_{\omega-\Omega}$~\footnote{The notation for $E$ is borrowed from the algebraically analogous Susskind-Glogower operators of quantum optics~\cite{susskind:1964}.}. These modes evolve in time according to their eigenvalue $\omega$, $\mathbf{a}_\omega(t) = \mathbf{a}_\omega e^{i \omega t}$. 
In general, $E$ can also apply a reciprocal phase shift ($\theta$ in Fig.~\ref{fig:fig1}b), which is hereafter set to $\theta=0$.
Even by itself, a single ambidirectional modulator can be non-reciprocal; time-reversal maps the down-converter onto an up-converter, $\Omega \rightarrow -\Omega$, in the same way that time-reversal alters magnetic fields, $B \rightarrow -B$, in ferrite circulators that operate via the Faraday effect.
Preserving the frequency of the input signal, however, requires that a down-converter be paired with an up-converter with transfer operator $E^\dagger \mathbf{a}_{\omega}~=~\mathbf{a}_{\omega + \Omega}$. Cascading complementary converters, though, makes a reciprocal network, as $ E E^{\dagger} - E^{\dagger} E = [E,E^\dagger] = 0$.

Non-reciprocity can be restored by inserting a time delay between the two frequency converters.  The transfer operator for a delay, $D$, translates a phasor $\mathbf{a}_{\omega}(t)$ forward in time by a time $\tau$ such that $D \mathbf{a}_{\omega}(t) = \mathbf{a}_{\omega}(t + \tau)$.  As time and frequency are Fourier duals, consecutive translations in these variables do not generally commute. 
Forward and reverse scattering parameters for the ladder gyrator 
are eigenvalues of the products of transfer operators, $E^\dagger D E \mathbf{a}_\omega = s_{12} \mathbf{a}_\omega$ and $E D E^\dagger \mathbf{a}_\omega = s_{21} \mathbf{a}_\omega$.  Here the scattering parameter $s_{nm}$ is a complex number describing the phase shift and the fractional change in amplitude of a signal which has propagated from port $m$ to port $n$. When $E$ and $D$ do not commute, the network is non-reciprocal:
\begin{equation}
\left(s_{12} - s_{21}\right)\mathbf{a}_\omega =  \left( E^{\dagger}[D,E] + [D,E]E^{\dagger} \right) \mathbf{a}_\omega.
\label{SMatrixCommutator}
\end{equation}
Evaluation of the commutator (see supplement) shows
\begin{equation}
|s_{21} - s_{12}| \propto \sin(\Omega \tau).
\label{laddercommutator}
\end{equation}
When $\Omega \tau = \pi/2$, non-reciprocity is maximized and fields flowing forward through the network are advanced in phase by $\pi$ relative to fields flowing backwards through the network.  


The convert-delay scheme can alternatively be realized using frequency-symmetric modulators, with transfer operators $X = \frac{1}{2} \left( E^{\dagger} + E \right)$ and $Y = \frac{1}{2 i} \left( E^{\dagger} - E \right)$.  Euler's formula gives these operators a simple form in the time domain, where their action corresponds to multiplication by a sinusoid (Fig.~\ref{fig:fig1}e). 

For these frequency-symmetric modulators, time-reversal alters the phase rather than the frequency of modulation. 
Nevertheless, non-commutation with time delay can still be used to construct a gyrator, as depicted in Fig.~\ref{fig:fig1}e.
In this design, the convert-delay-convert protocol is duplicated in a parallel arm of the circuit, with the phases of the frequency modulators in the second arm advanced by $\pi/2$ to coherently erase the second harmonics of the modulation frequency. 
The difference between the forward and backward scattering parameters $s_{21} - s_{12}$ of such a device is again a commutator of transfer operators (full calculation shown in supplementary information), which is also described by Eq.~\ref{laddercommutator}.  As in the ladder gyrator, gyration is realized when $\Omega \tau = \pi/2$.  
We call this gyrator the balanced gyrator, for the 
way in which it arranges the destructive interference of second harmonics.

Critically, the non-reciprocal phase acquired in both the ladder gyrator and the balanced gyrator depends only on the length of the delay and the frequency of modulation, and is independent of the signal frequency. This endows the devices with a broad bandwidth, which is limited only by the bandwidths of their constituent elements. We emphasize that this entire bandwidth can be utilized simultaneously due to the linear relationship between input and output signals in all device components; a single wave-packet, with a spectral width of several GHz, will gyrate in both devices.  

\begin{figure}[t] 
\begin{center}
\includegraphics[width=1.0\linewidth]{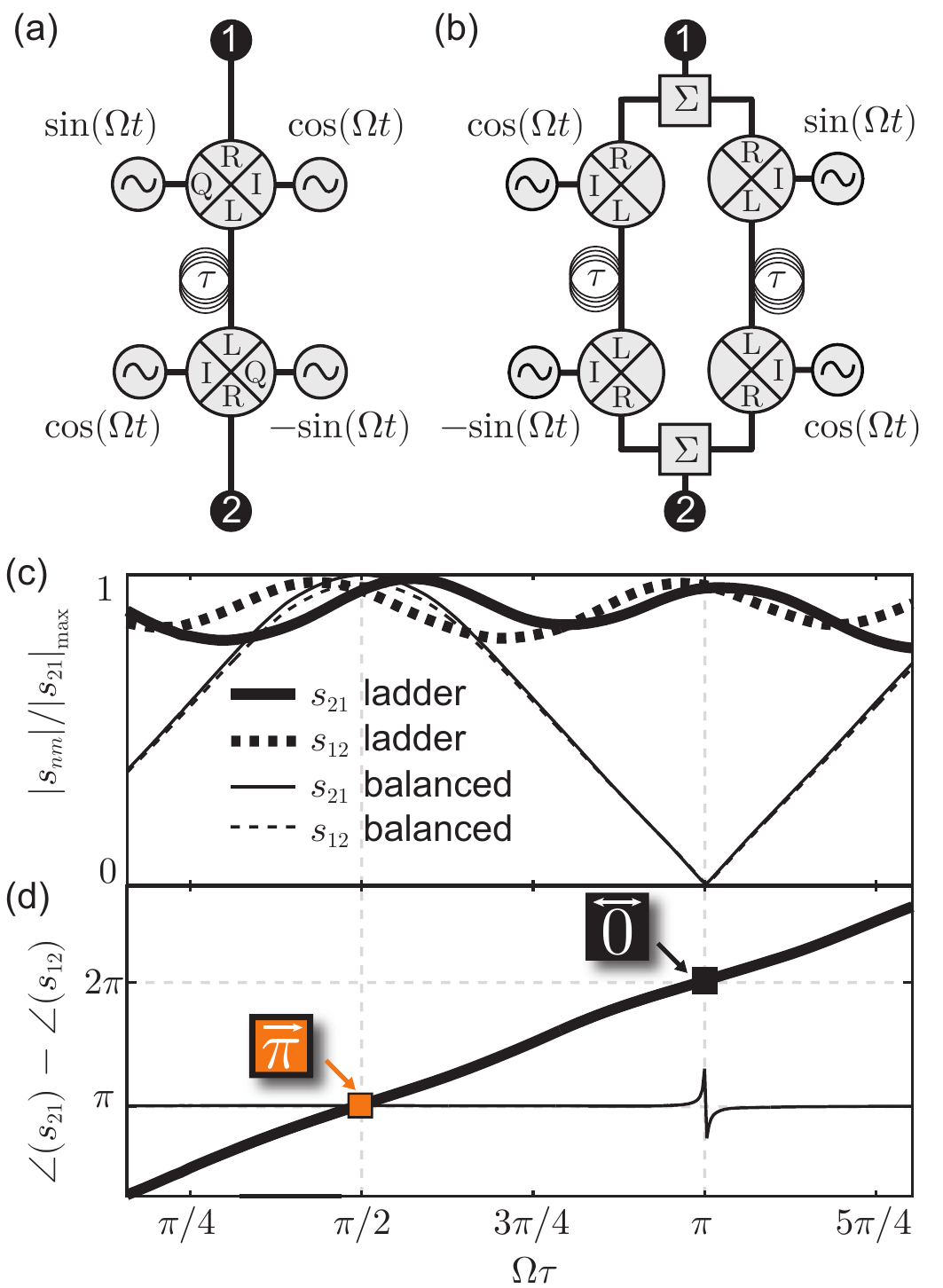}
\caption {Circuit schematics of a ladder gyrator (a) and a balanced gyrator (b) realized with IQ mixers (crossed circles) 
as the frequency modulation elements and coaxial transmission-line cables (coil symbols) for delay $\tau$. Signals in the balanced gyrator are split into two paths and recombined using reciprocal summation elements (labeled $\Sigma$).
In an IQ mixer, bias tones $I(t)$ and $Q(t)$ multiply the incident signal by $I(t) - i Q(t)$ leading to frequency conversion of fields moving between ports L and R. In the balanced gyrator, $Q(t) = 0$.
(c) The measured device transmission, normalized to the maximum forward transmission of each device $|s_{21}|_{\mathrm{max}}$ and (d) the measured phase difference are plotted for both circuits, as a function of the relative phase $\Omega \tau$ acquired in the delay.
Orange and black boxes indicate operation points where the ladder circuit acts as a gyrator or as a reciprocal element, which we call a `reciprocator'.
}
\label{fig:fig2}
\end{center}
\end{figure}

\begin{figure}[htb]
\begin{center}
\includegraphics[width=1.0\linewidth]{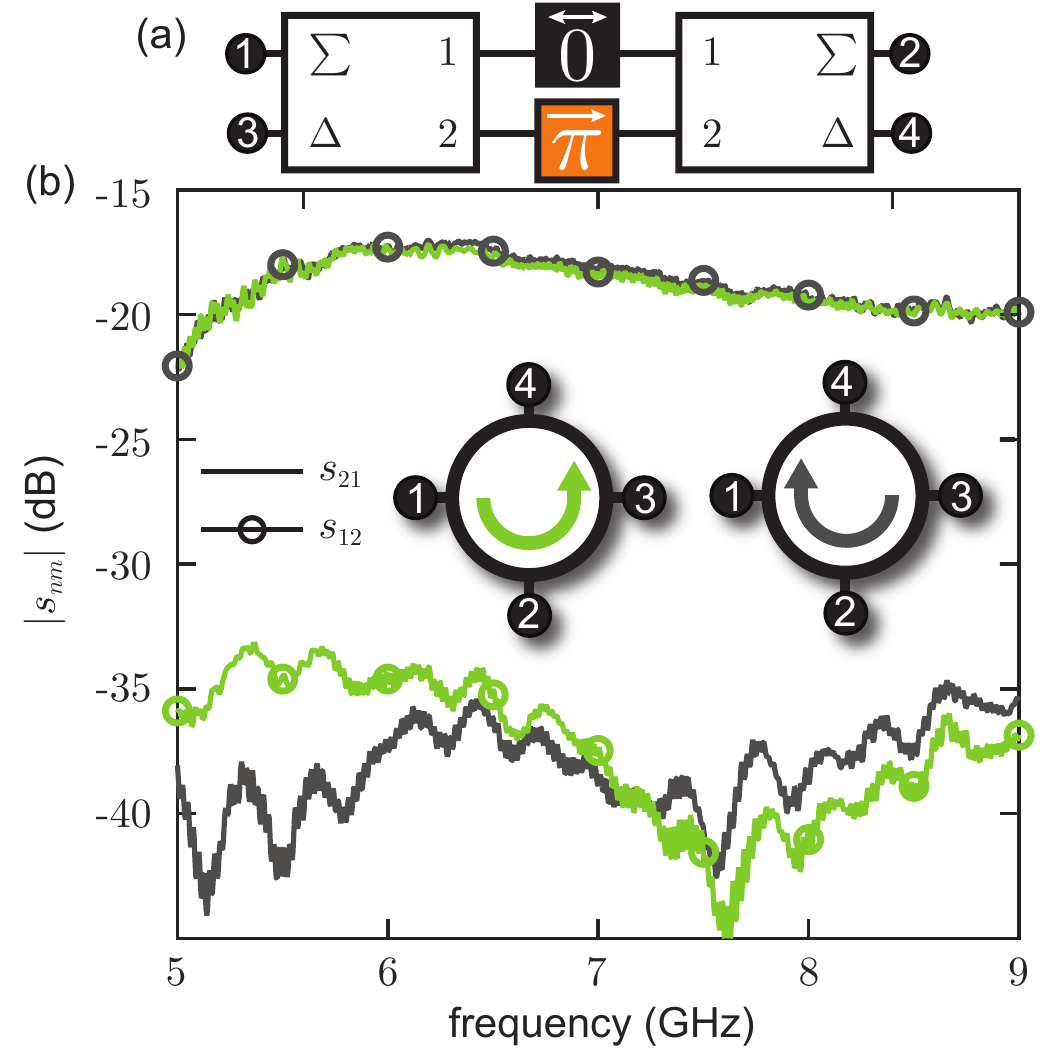}
\caption {
(a) A circulator is constructed from a ladder gyrator and a reciprocator placed between two microwave beam splitters: linear, reciprocal elements that transform signals incident on ports $1$ and $2$ into their sum and difference, exiting at ports $\Sigma$ and $\Delta$ respectively. 
(b) Forward (solid lines) and reverse transmission (solid lines with circles) between two adjacent ports of the circuit, when the device is configured as a clockwise circulator (gray traces) and a counter-clockwise circulator (green traces).
Relative isolation greater than 15 dB is achieved between 5 and 9 GHz. A moving average filter (span 15 MHz) has been applied to the isolated traces to smooth a several-dB ripple resulting from reflections within the delay lines.
}
\label{fig:fig3}
\end{center}
\end{figure}

\textit{Gyrator measurements.---} 
Measuring the phase difference acquired by a forward propagating signal compared to the reverse signal reveals the astounding bandwidth of the gyrators. Fig.~\ref{fig:fig1}f shows the non-reciprocal phase shift $\angle s_{21}-\angle s_{12}$ for a ladder gyrator (orange trace) and a balanced gyrator (green trace). The mean value of this phase difference, measured between 4 to 10 GHz, is $180.05 \degree \pm 1.46 \degree$ for the ladder gyrator and $180.41 \degree \pm 3.72\degree$ for the balanced gyrator.
The devices are constructed with mixers (acting as frequency modulation elements) and 
coaxial transmission-line cables (to create delay). Experimental schematics of the circuits are shown in Fig.~\ref{fig:fig2}a~and~b, and photographs of the circuits are shown in Fig.~S1a~and~b.

The performance of both gyrators hinges on correctly setting the dynamical phase $\Omega \tau$ acquired in the delay lines. To show this, the amplitude of the forward and reverse scattering parameters for both devices is plotted in Fig.~\ref{fig:fig2}c, as a function of the delay phase $\Omega \tau$. Experimentally, this parameter is tuned by varying the modulation frequency between 4 and 33 MHz. In the balanced gyrator, the amplitude of transmission is reciprocal with a magnitude that depends on the delay phase.
This magnitude is maximized at the operation point $\Omega \tau = \pi/2$, and minimized when up and down converted signals interfere destructively at $\Omega \tau = \pi$, which is quantitatively consistent with expectations.
In contrast, transmission in the ladder gyrator is nearly independent of delay phase. Observed non-reciprocal ripples in the magnitude of transmission are attributed to spurious sidebands created by the mixers, which create alternative frequency-pathways for transmission and interfere non-reciprocally with the intended pathways shown in Fig.~\ref{fig:fig1}c. In the balanced gyrator, this effect is suppressed by the symmetry of the device and, as a consequence, $s_{21}$ and $s_{12}$ are close in magnitude for all values of $\Omega \tau$. 

Fig.~\ref{fig:fig2}d shows the phase difference $\angle s_{21} - \angle s_{12}$ as a function of $\Omega \tau$. In the balanced gyrator, $\angle s_{21} - \angle s_{12}$ is nearly independent of $\Omega \tau$, except near $\Omega \tau = \pi$, where transmission vanishes.  Fitting the data to a constant function yields $\angle s_{21} - \angle s_{12} = 180.8^\circ \pm 5.3^\circ$. In the ladder gyrator, however, the phase difference depends strongly on $\Omega \tau$.  We observe a linear relationship between $\angle s_{21} - \angle s_{12}$ and $\Omega \tau$ and extract a slope of $2.010 \pm 0.003$, close to the expected value  $( \angle s_{21} - \angle s_{12} ) / \Omega \tau = 2$. This dependence on delay phase is a useful property, as it allows the ladder circuit to be operated as a gyrator when $\Omega \tau = \pi/2$ (orange gyrator symbol in Fig.~\ref{fig:fig2}d), or as a reciprocal device with the same frequency-dependent attenuation as the gyrator, when $\Omega \tau = \pi$ (black reciprocal symbol in Fig.~\ref{fig:fig2}d).  When operated in this reciprocal way, we refer to the ladder circuit as a `reciprocator'. 

The fact that the non-reciprocity in both gyrators depends identically on the delay phase $\Omega \tau$ (Eq.~\ref{laddercommutator}) suggests a similarity in their operation.  
Interestingly, Fig.~\ref{fig:fig2} shows that this equivalence is achieved in different ways.
The ladder gyrator exhibits roughly delay-independent transmission, but delay-dependent non-reciprocal phase.
The balanced gyrator exhibits delay-independent non-reciprocal phase, but delay-dependent transmission.
This amplitude-phase duality reflects the fact that the two gyrators are interferometric duals, in the sense that the balanced gyrator is equivalent to an interferometer with ladder gyrators in each arm.
As interferometers translate phase difference into transmission amplitude, the two gyrators necessarily exhibit a dual amplitude-phase relationship.

Finally, to ensure the convert-delay approach does not create spurious sidebands, a spectrum analyzer is used to measure the power transmitted at harmonics of the modulation frequency $\Omega$.  
Sideband suppression is especially important to the performance of a broadband gyrator, in which harmonics may still lie within the device's operation bandwidth.  Fig. S3 in the supplementary information shows the power transmitted through both devices at even harmonics of the modulation frequency. (Odd harmonics are suppressed by symmetry). At the devices' operation points, spurious harmonics are suppressed by 15 dB in both gyrators.

\textit{Circulator.---}
To demonstrate the utility of a broadband gyrator, we use a ladder gyrator to create a broadband four-port circulator using a Hogan-like construction~\cite{hogan:1953}.  Fig.~\ref{fig:fig3}a shows a schematic of the circulator, made from a ladder gyrator and a reciprocator connecting two 
microwave beam splitters (circuit photograph in Fig.~S1c).  
Using a reciprocator (instead of a simple through) ensures equality of the electrical length and the frequency-dependent insertion loss in each path between the hybrids. It also prevents transmission between isolated ports outside of the circulator's operation band.

The direction of circulation may be reconfigured in-situ either by changing the modulation frequencies of the two devices (exchanging the reciprocator and the gyrator) or inverting the phases of the modulation tones (reversing the gyration direction). Fig.~\ref{fig:fig3}b shows circulator scattering parameters $s_{21}$ and $s_{12}$, when the device is configured for both clockwise and counterclockwise circulation. Greater than 15 dB of relative isolation is achieved between 5 and 9 GHz. The full 16 element scattering matrix of the device is shown in~Fig.~S4, where reverse isolation may be as great as 30 dB over a 400 MHz bandwidth.


In addition to isolation and bandwidth, insertion loss is another important metric for circulator performance.  The circulator shown in Fig.~\ref{fig:fig3} has a large insertion loss of approximately $20$ dB, primarily due to the attenuation of the delays ($\approx 5$ dB) and the conversion gain of the commercial mixers ($\approx -6$ dB). These sources of loss, however, are not fundamental to the approach. 
A related construction was recently demonstrated on-chip for superconducting circuits with the delay lines replaced by frequency-tunable resonances, allowing lossless non-reciprocity over a restricted bandwidth~\cite{kerckhoff:2015,chapman:2017b}. Ambidirectional modulators using purely reactive elements have also recently been developed \cite{chapman:2016,chapman:2017,naaman:2017}.
A completely lossless, broadband superconducting implementation appears to be possible in principle, although an explicit construction remains an open problem.

Our concept is likely to be applied first in situations where large bandwidth, device integration and device miniaturization are imperative. Applications include, optical and wireless telecommunications and emerging millimeter wave sensing technology. A monolithic integrated-circuit implementation (as in Ref.~\cite{reiskarimian:2016}) is particularly appealing for microwave and millimeter wave telecommunication applications, for example, antenna duplexing in small cell base-stations and mobile phones. Such a device would use microwave frequency bias tones, thus allowing for only centimeter length delay lines, which can be integrated onto a chip less than $1\times1$ cm$^2$ in size. Greater integration and miniaturization could be achieved by synthesizing the necessary bias tones on the same chip. It should also be possible to realize integrated optical nonreciprocal devices using our concept because silicon photonic devices provide both low loss on-chip delay elements \cite{cardenas:2009} and high speed optical modulators whose performance is rapidly improving \cite{reedGT:2010,sun:2016}.

\textit{Conclusion.---} We present two gyrator designs, both of which take a convert-delay approach for generating non-reciprocity, founded on the non-commutation of successive translations in time and frequency. Both devices operate over an octave bandwidth, and can be used to build other broadband nonreciprocal circuits. Broadband performance is demonstrated explicitly by constructing a non-magnetic microwave circulator with over 15 dB of relative isolation across the 5 to 9 GHz band. This approach provides a general prescription for generating broadband non-reciprocity, applicable in any platform that contains frequency converting elements, including on-chip implementations for quantum information processing, nanophotonics, or telecommunication.

\vspace{0.1in}
This work is supported by the ARO under contract W911NF-14-1-0079 and the National Science Foundation under Grant Number
1125844.


%

\pagebreak
\widetext
\begin{center}
\textbf{\large Supplementary Material for\\ ``Breaking Lorentz reciprocity with frequency conversion and delay''}
\end{center}
\setcounter{equation}{0}
\setcounter{figure}{0}
\setcounter{table}{0}
\setcounter{page}{1}
\makeatletter
\renewcommand{\theequation}{S\arabic{equation}}
\renewcommand{\thefigure}{S\arabic{figure}}
\renewcommand{\bibnumfmt}[1]{[S#1]}
\renewcommand{\citenumfont}[1]{S#1}

\section{Experimental setup}
\begin{figure}[htb]
\begin{center}
\includegraphics[width=1.0\linewidth]{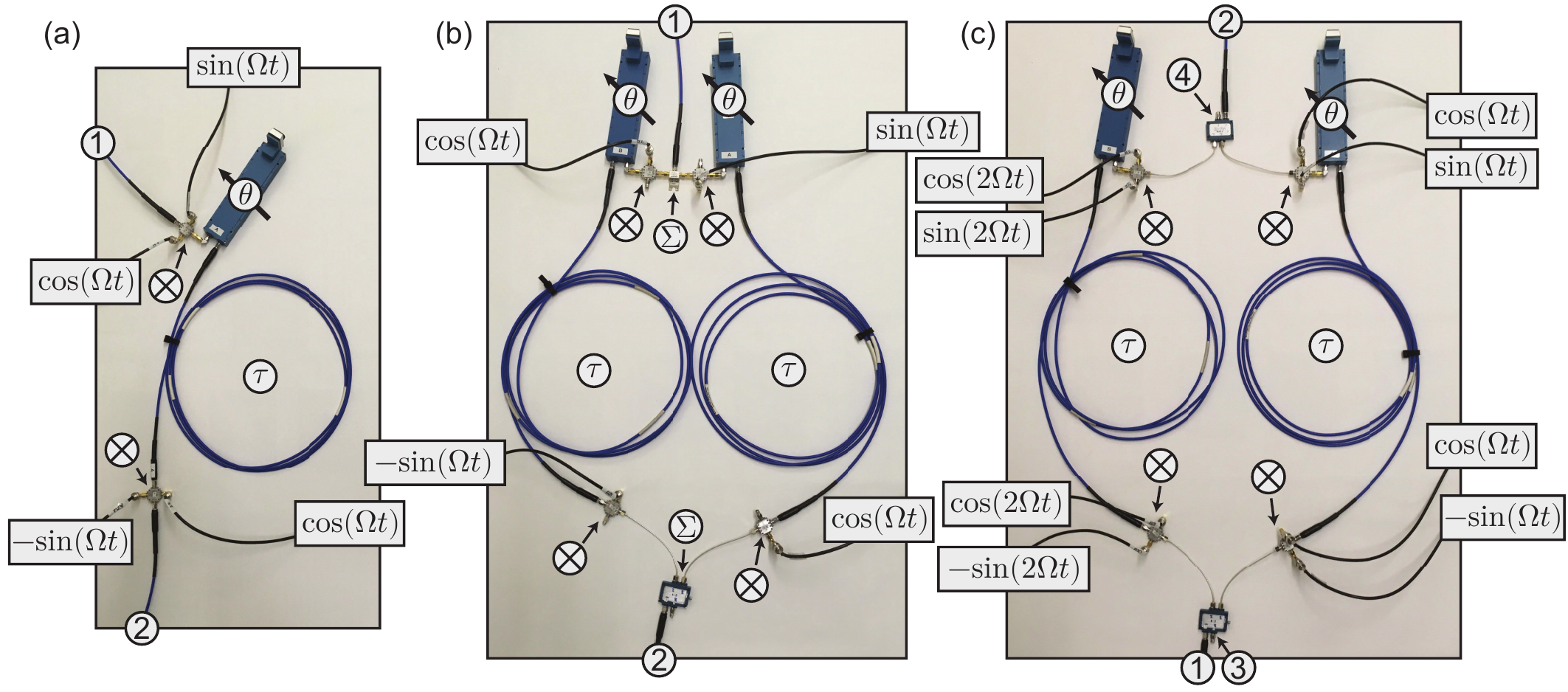}
\caption {\small Device photographs: (a) Ladder gyrator. (b) Balanced gyrator. (c) Circulator.}
\label{fig:supp1}
\end{center}
\end{figure}

Photographs of the ladder gyrator (Fig.~\ref{fig:supp1}a), the balanced gyrator (Fig.~\ref{fig:supp1}b) and the circulator (Fig.~\ref{fig:supp1}c) are shown. The devices are constructed with delay lines (SMA cables, Mini-Circuits 12 ft. SMSM+ 50 Ohm) and IQ mixers (Marki IQ-4509). Mixer I and Q ports are biased with a peak-to-peak voltage of 1.4 V. Double-balanced mixers in the balanced gyrator are implemented by terminating the Q port of the IQ mixers with 50 Ohms. 
Two microwave mechanical phase shifters (ARRA 6425E) are placed in each delay line to fine tune electrical length. 
180$\degree$ hybrids (Krytar 4040124) and Wilkinson power splitters (Mini-Circuits ZX10-2-98+) are used for power division and recombination. Precision microwave elbows (Pasternack PE91152) are included in each delay line for convenient assembly. 

\section{Gyrator scattering parameters}

\begin{figure*}[htb]
\begin{center}
\includegraphics[width=1.0\linewidth]{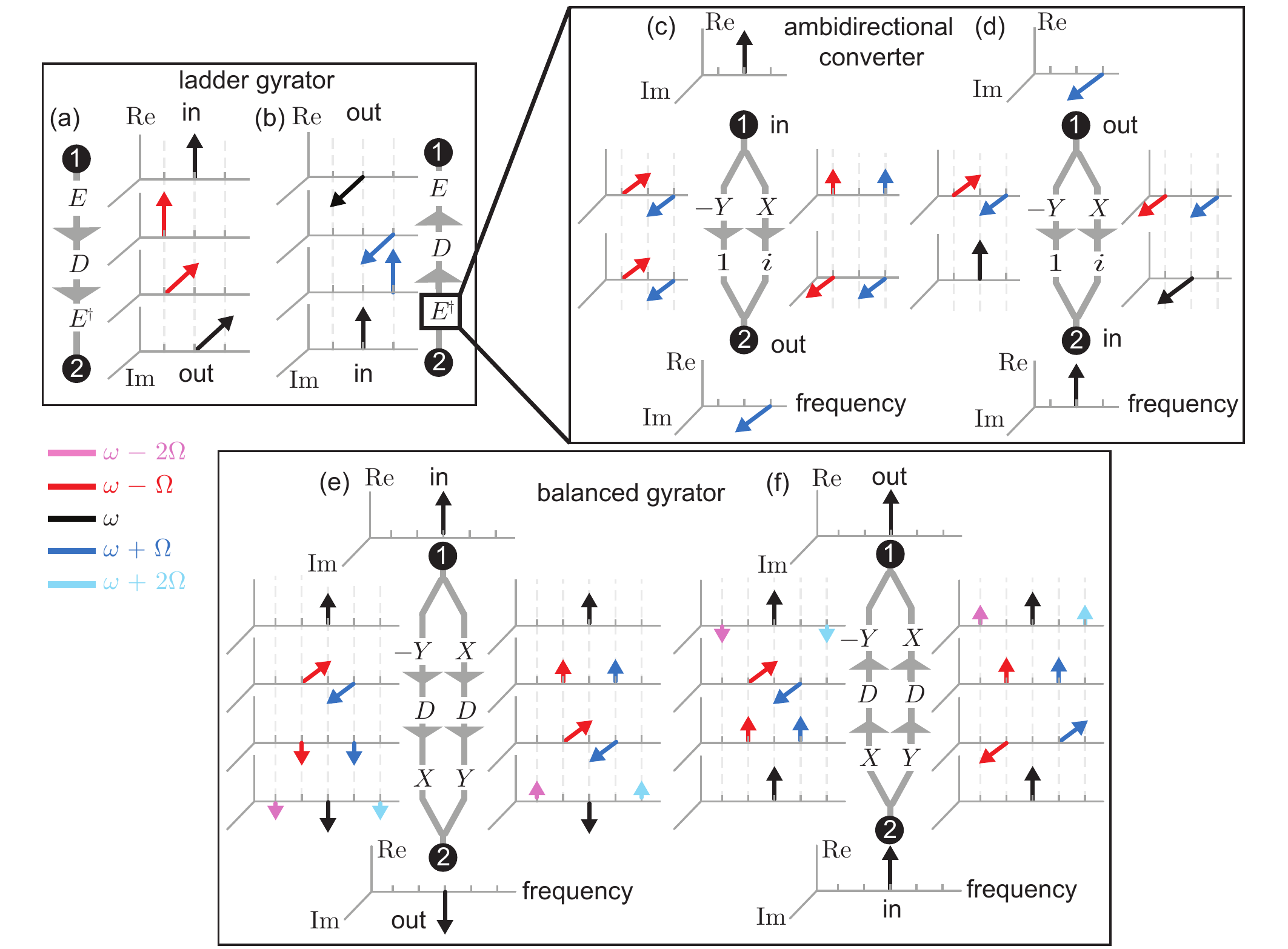}
\caption {
Ladder gyrator transmission in the forward (a) and reverse (b) direction represented in phasor notation. A ladder gyrator is composed of ambidirectional frequency converters implemented with IQ mixers: forward (c) and reverse (d) operation of an up-converter is described. Symbol $1$ corresponds to an identity operation and symbol $i$ corresponds to a $90 \degree$ phase shift of all itinerant signals (for more information and a schematic see Ref.~[2]). Balanced gyrator transmission in the (e) forward and (f) reverse direction.
}
\label{fig:supp2}
\end{center}
\end{figure*}

The scattering matrix for each gyrator is calculated using transfer matrix operators for delay and frequency conversion. 
Acting on an itinerant mode $\mathbf{a}_\omega$ which evolves in time according to its eigenvalue $\omega$, $\mathbf{a}_\omega(t) = \mathbf{a}_\omega e^{i \omega t}$, these operators are described by translations in time or in frequency,
\begin{subequations}
\begin{align}
D \mathbf{a}_\omega (t) = \mathbf{a}_\omega (t + \tau), \\
E \mathbf{a}_\omega = \mathbf{a}_{\omega - \Omega}.
\end{align}
\label{TransferOperators}
\end{subequations}
Up-conversion is the inverse of down conversion: $E^{\dagger} \mathbf{a}_\omega = \mathbf{a}_{\omega + \Omega}$. We implement these operations with delay lines of length $\tau c$ ($c$ the speed of light in a coaxial cable), and IQ mixers with I and Q ports biased in quadrature at frequency $\Omega$.

Down-conversion, delay and up-conversion are cascaded in series to create the ladder device. Forward and reverse transmission (Figs.~\ref{fig:supp2}a and \ref{fig:supp2}b), respectively, are given by,
\begin{subequations}
\begin{align}
E^{\dagger} D E \mathbf{a}_\omega  = e^{i (\omega-\Omega) \tau} \mathbf{a}_\omega, \\
E D E^{\dagger} \mathbf{a}_\omega  = e^{i (\omega + \Omega) \tau} \mathbf{a}_\omega. 
\end{align}
\label{UDGyratorSParams}
\end{subequations}
The full scattering matrix can be calculated directly using these transfer operators: off-diagonal scattering matrix elements are the eigenvalues of forward and reverse transmission (Eq.~\ref{UDGyratorSParams}). Diagonal scattering matrix elements are zero when circuit components are impedance matched. The scattering matrix for the ladder device therefore becomes,
\begin{equation}
e^{i\omega \tau} \left(\begin{array}{cc} 0 & e^{+ i \Omega \tau}\\ e^{- i \Omega \tau} & 0 \end{array}\right).
\label{UDGyratorSMatrix}
\end{equation}
Transmission magnitude is independent of $\Omega \tau$, but the phase difference between forward and reverse transmission is maximized for $\Omega \tau = \pi/2$. The difference between off-diagonal scattering matrix elements is,
\begin{equation}
s_{12} - s_{21} = 2 i e^{i\omega \tau} \sin(\Omega \tau).
\label{UDGyratorSParamsDiff}
\end{equation}

In contrast to the ladder gyrator, the balanced gyrator is described by  
linear combinations of the frequency conversion operators,
\begin{subequations}
\begin{align}
X = \frac{1}{2} \left( E^{\dagger} + E \right),  \\
Y = \frac{1}{2 i} \left( E^{\dagger} - E \right).
\label{CSoperators}
\end{align}
\end{subequations}
To calculate transmission through the balanced device, it is necessary to sum the transmission through each of its two arms. Each arm by itself is non-reciprocal at the carrier frequency, but creates reciprocal second harmonics of the modulation frequency, which are canceled upon recombination. Forward and reverse transmission (Figs.~\ref{fig:supp2}e and \ref{fig:supp2}f) are given by the following two equations, respectively,
\begin{subequations}
\begin{align}
\frac{1}{2} \left(X D (-Y) + Y D X \right) \mathbf{a}_\omega = - \frac{1}{2} e^{i\omega \tau} \sin(\Omega \tau) \mathbf{a}_\omega, \\   
\frac{1}{2} \left((-Y) D X + X D Y \right) \mathbf{a}_\omega = + \frac{1}{2} e^{i\omega \tau} \sin(\Omega \tau)   \mathbf{a}_\omega.
\end{align}
\label{CSGyratorSParams}
\end{subequations}
As before, the scattering matrix off-diagonal elements are the eigenvalues of transfer operators (Eq.~\ref{CSGyratorSParams}). The scattering matrix for the balanced device is,
\begin{equation}
\frac{1}{2}e^{i\omega \tau}\left(\begin{array}{cc} 0 & \sin(\Omega \tau) \\ -\sin(\Omega \tau) & 0 \end{array}\right).
\label{CSGyratorSMatrix}
\end{equation}
In contrast to the ladder device, the magnitude of transmission is proportional to $\sin(\Omega \tau)$, but forward and reverse transmission are always $\pi$ out of phase,
\begin{equation}
s_{12} - s_{21} = e^{i\omega \tau} \sin(\Omega \tau).
\label{CSGyratorSParamsDiff}
\end{equation}
Consequently, non-reciprocity is again maximized when $\Omega \tau = \pi/2$.

It is noted that both the ladder gyrator and balanced gyrator split incident signals into multiple paths to coherently cancel sidebands. In the ladder gyrator, this cancellation occurs inside of each IQ mixer, as illustrated in Fig.~\ref{fig:supp2}c and d. 
In the balanced gyrator, second harmonics are cancelled when both paths of the device are recombined.

\section{Gyrator sidebands}

\begin{figure}[htb]
\begin{center}
\includegraphics[width=1.0\linewidth]{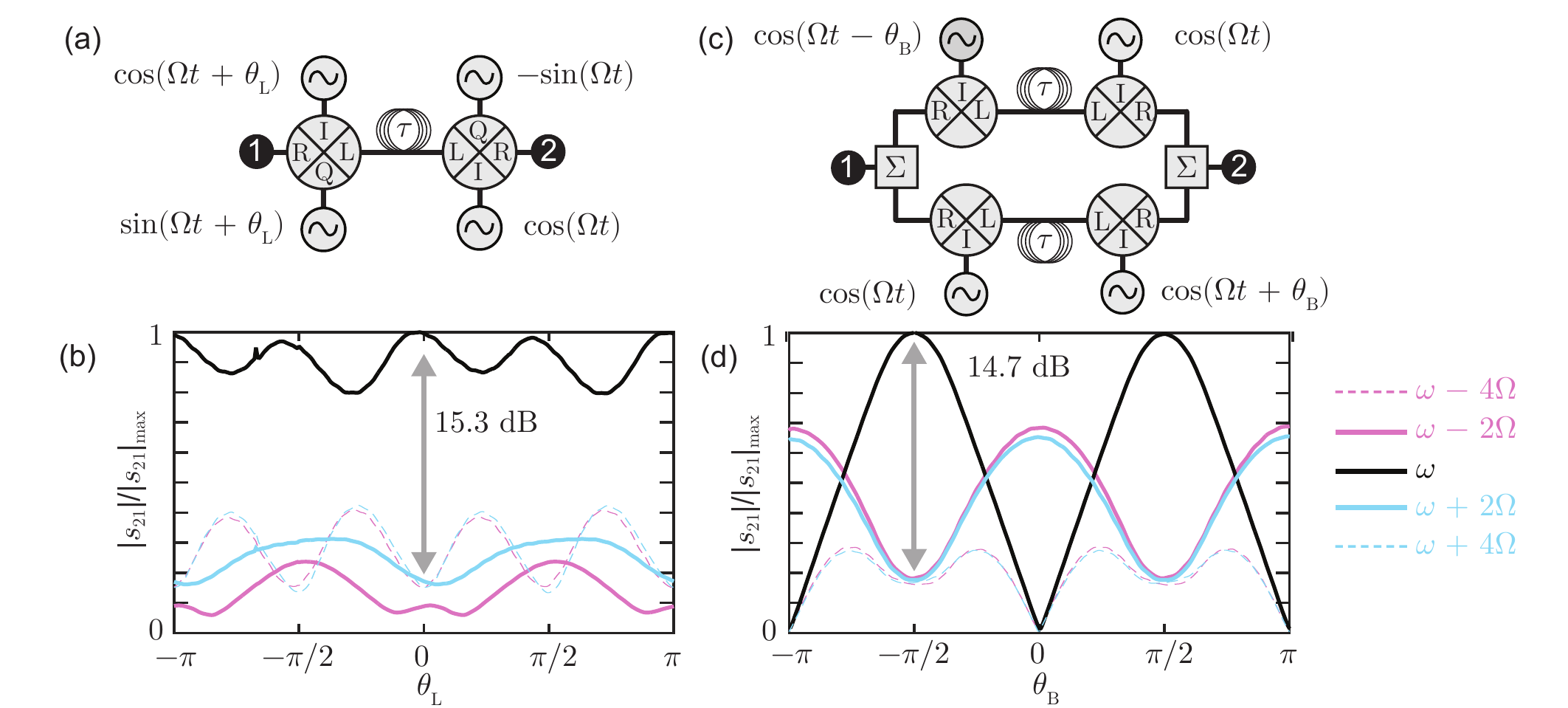}
\caption {\small (a) Schematic of the ladder gyrator. $\theta_L$ is the relative phase difference between bias tones on the different modulators. (b) Measured ladder gyrator sidebands are plotted as a function of $\theta_L$. Sideband suppression of 15.3 dB occurs at $\theta_L = 0$. (c) Schematic of the balanced gyrator. A phase difference $\pm \theta_{B}$ is swept along a pair of diametrically opposite bias lines. (d) Measured balanced gyrator sidebands are plotted as a function of $\theta_{B}$. Sideband suppression of 14.7 dB occurs at $\theta_{B} = \pm \pi/2$. At $\theta_{B} = 0$, power is split evenly between second harmonics.}
\label{fig:supp3}
\end{center}
\end{figure}

Heavy suppression of all transmitted sidebands is an important feature of all the devices presented in this Letter. 
To show this, a transmission spectrum is measured for both gyrators for incident signals at $\omega/2\pi$ = 5.5 GHz and a modulation frequency of $\Omega/2\pi$ = 12.665 MHz.  For the ladder gyrator, transmitted spectral components at the input frequency and the four largest sidebands are plotted as a function of $\theta_L$, the relative phase difference between the bias tones of the up-converting modulator and the down-converting modulator (Fig.~\ref{fig:supp3}a). At $\theta_L = 0$, sidebands are suppressed by 15.3 dB (Fig.~\ref{fig:supp3}b). Likewise, sidebands of the balanced gyrator are measured as a function of $\theta_B$, (defined in Fig.~\ref{fig:supp3}c). At $\theta_B = 0$, sidebands are suppressed by 14.7 dB or greater (Fig.~\ref{fig:supp3}d).  In both devices, sideband suppression is limited by the generation of spurious sidebands in the IQ mixers.




\section{Circulator scattering matrix}

\begin{figure}[htb]
\begin{center}
\includegraphics[width=1.0\linewidth]{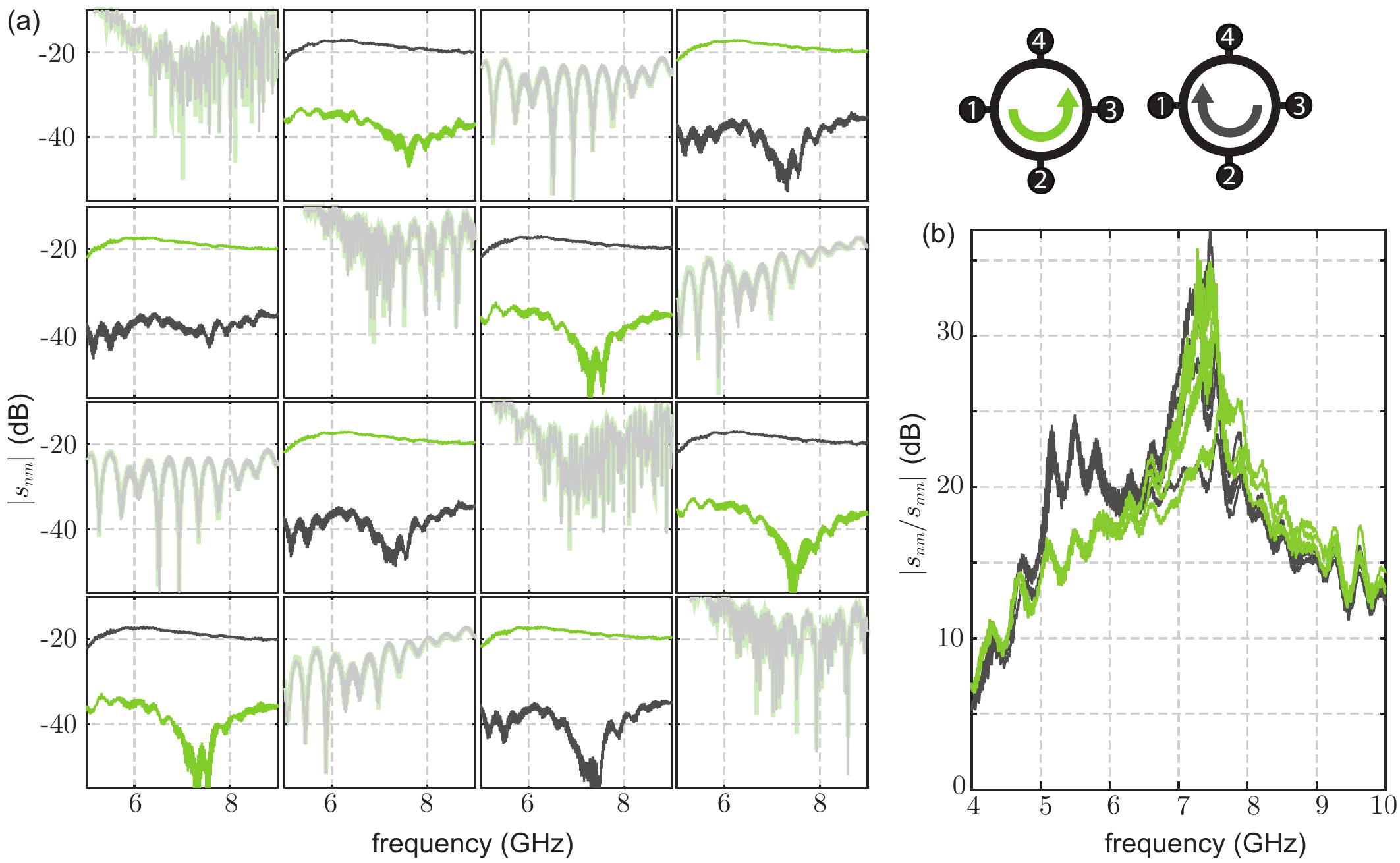}
\caption {\small (a) Scattering matrix for the four port circulator. (b) Contrast between non-reciprocal transmissive and isolated scattering elements $s_{nm}/s_{mn}$. All four combinations are plotted for each direction of circulation. A moving average filter (span 15 MHz) has been applied to the isolated traces before computing the contrast.}
\label{fig:supp4}
\end{center}
\end{figure}

The full scattering matrix of the four-port broadband microwave circulator is shown in Fig.~\ref{fig:supp4}a. Non-reciprocal matrix elements (bold colors) show transmission and reverse isolation. Diagonal matrix elements (faded colors) quantify the impedance match of the device, and reciprocal off-diagonal elements (also faded colors) quantify the reverse isolation of the commercial hybrid couplers. The contrast between non-reciprocal matrix elements $s_{nm}/s_{mn}$ is plotted in Fig.~\ref{fig:supp4}b. Contrast is greater than 10 dB over greater than an octave of bandwidth, and can be greater than 30 dB over as much as 400 MHz. At 6 GHz, the scattering matrix (in dB) for the device operated as a counter-clockwise circulator (green) is,
\begin{equation}
    \left(\begin{array}{cccc} 
    -19.43 & -34.40 & -27.33 & -17.40 \\
    -17.35 & -13.92 & -34.54 & -24.78 \\
    -27.33 & -17.40 & -16.27 & -34.33 \\
    -34.48 & -24.82 & -17.30 & -13.33
\end{array}\right).
\label{GyratorSMatrix}
\end{equation}

Insertion loss of the circulator is between 17.11 dB to 17.40 dB when operated at 6 GHz. Loss comes from the following sources: (1) between -5.96 dB to -6.59 dB of up-conversion gain, and -5.50 dB to -5.93 dB of down-conversion gain, in each of the four component single sideband modulators. (2) 4.34 dB and 4.86 dB of insertion loss in the two delay lines (including microwave phase shifters). (3) between 0.571 dB to 0.74 dB of insertion loss in each of the 180$\degree$ hybrid couplers (including 15 cm SMA cables on ports one and two of the hybrids). 


\end{document}